\documentclass[epj]{svjour}
\usepackage{latexsym}
\usepackage{graphics}

\begin{document}

\newcommand{\ket}[1]{\left| #1 \right\rangle}
\newcommand{\me}[3]{\left\langle #1\left| #2\right| #3\right\rangle}
\newcommand{\bra}[1]{\left\langle #1 \right|}

\title{Non-localities in nucleon-nucleus potentials}
\subtitle{and their effects in nucleon-nucleus scattering}
\author{P. Fraser 
\inst{1}
\and
K. Amos 
\inst{1}
\and 
S. Karataglidis
\inst{2}
\and
L. Canton
\inst{3}
\and
G. Pisent
\inst{3}
\and
J. P. Svenne
\inst{4}
}                     

\institute{
School of Physics, University of Melbourne,
Melbourne, Victoria 3010, Australia
\and
Department of Physics and Electronics, Rhodes University,
Grahamstown 6140, South Africa
\and
Istituto  Nazionale  di  Fisica  Nucleare,
   Sezione  di Padova,\\
 e  Dipartimento di Fisica  dell'Universit\`a
   di Padova, Padova I-35131, Italia
\and 
Department  of  Physics  and Astronomy, University  of Manitoba,\\
 and Winnipeg  Institute for Theoretical Physics,
 Winnipeg, Manitoba R3T 2N2, Canada}

\date{Received: date / Revised version: date}

\abstract{
Two causes of non-locality inherent in  nucleon-nucleus scattering
are considered. They are the results of
two-nucleon antisymmetry of the projectile with each nucleon in the
nucleus  and  the  dynamic polarization potential representation of
channel coupling.   For energies $\sim 40 - 300$ MeV, a $g$-folding
model of the optical potential is used to show the influence of the
knock-out process that is a result of the two-nucleon antisymmetry.
To explore  the  dynamic  polarization  potential caused by channel
coupling, a multichannel algebraic scattering  model  has been used
for low-energy scattering.}


\maketitle

\section{Introduction}
\label{intro}

Non-locality is omnipresent in modern physics having relevance to a quite 
diverse array of topics. Examples are the heavy-element abundance patterns in 
white dwarf stars, the bulk modulus of silicon, the tensionless limit of 
string theory,
non-commutative quantum field theory, image retrieval in scanning probe 
microscopy, electron crystallography, Bose-Einstein condensates, quantum
nanostructures, high Reynolds number flow and optical recognition of musical 
scores.

In nuclear physics, the nucleon-based structure of the nucleus and the 
scattering
and reactions of a projectile (nucleon or nucleus) with  a target nucleus,  are
complicated many-body problems. They involve non-local
interactions, and one of the major sources of 
non-locality arises from the effects of the Pauli principle. The many-nucleon 
wave function of the system must be antisymmetric with respect to interchange
of all pairs of nucleons. This
leads to the interaction of an individual nucleon with the rest of the system 
(the ``optical potential'' in the case of scattering) having 
both direct and exchange (knock-out) components. The exchange components are 
intrinsically 
non-local. Because of the short-ranged nature of the strong
nucleon-nucleon ($NN$) interaction, the exchange amplitudes have significant 
magnitudes; ones comparable to direct scattering amplitudes.

In this paper,
we deal specifically with non-localities arising in effective interaction 
methods of modelling elastic nucleon-nucleus ($NA$) scattering.  The optical 
potential resulting from such theories is complex (when there is flux loss 
from the incident channel), energy dependent, and non-local. These
characteristics are due in part to 
the complex nature of the effective $NN$ interaction underlying the
process being  modified
from the free $NN$ force because of the presence of the nuclear
medium and from effects of the Pauli principle~\cite{Am00}.  A non-local and
complex optical potential results also 
from truncating the Hilbert space~\cite{Fe92}. 

The most common approach to analyze (elastic) scattering data is
the phenomenological optical model (OM) in which, usually, the $NA$ interaction
potential is chosen to have Woods-Saxon (WS) form with complex strengths, 
and central plus spin-orbit terms. Those optical potentials are local in form
with the energy dependence of the parameters assumed to account for non-local
effects.  One exception is the specific energy-independent, non-local potential 
studied by Perey and Buck~\cite{Pe62}.  
There have been other non-local $NA$ OM potentials proposed, 
but those invariably have
parameters to be specified in searches for good fits to each specific 
elastic-scattering data set. 
Such phenomenological approaches, and particularly wave functions 
derived therefrom, are not reliable. The actual non-localities
resulting from more precise treatment of Pauli principle effects, such
as with the exchange (knock-out) amplitudes in scattering, and/or of specific 
accounting for channel-coupling processes, lead to specific energy dependent
non-locality effects that are comparable with the non-exchange
amplitudes and/or local form corrections to a local optical potential 
for the channel-coupling processes. In part, the problems are hidden by
parameter adjustments in the approximate models being able to give 
asymptotic phase shifts which lead to reasonable fits to cross-section
and other data.  
Of course, there are a number of ways to determine a local equivalent
potential given a specific non-local one. A review of these forms
Chapter 11 in Ref.~\cite{Am00}.
The results usually are energy dependent and usually involve differentials.
However, to fit data with these potentials still requires  determination of the 
parameter set involved. A significant effect of the
violation of the Pauli principle inherent may persist~\cite{Ca05}. 
Nonetheless a major problem with use of any local equivalent form 
lies with just how large the
off-diagonal properties of the non-local optical potentials are and/or how
large, and with what interference, are their contributions to scattering.

For different energy regimes, when one seeks to minimize phenomenology, 
different models of $NA$ scattering are relevant. At energies greater than
the excitation energies of giant resonances in the target, 
a $g$-folding model~\cite{Am00} built upon the 
Kerman-McManus-Thaler (KMT) theory of scattering~\cite{Ke59} define  
microscopic
optical potentials.  That approach uses complex, energy- and density-dependent 
$NN$ effective interactions.  With optical potentials built using
the Melbourne force~\cite{Am00} as the effective interaction,
nucleon-nucleus scattering over the entire mass range and for energies
$\sim 40$ to over 200 MeV has been analyzed successfully~\cite{Am00}.
In the regime of energies coinciding with the excitation 
values of giant resonances, one needs consider their virtual excitation  
at least as a second-order exchange process~\cite{Ge75}. 
However, at energies from 0 to those for which the giant resonances are  
expected to be dominant doorway states, and when the target spectrum is
discrete, it is more appropriate to use a coupled-channel method 
to analyze scattering data. That is especially so whenever the discrete
spectrum of the target has low-lying collective states. 
However, most coupled-channel methods of solving $NA$ scattering problems 
are based on a collective-model formalism and limited to using local form 
factors.  These methods approximate non-localities with local effects by 
taking the parameters, used in finding solution of the coupled equations 
in coordinate space, to be energy dependent.  But in so doing the Pauli
principle is violated with serious effects~\cite{Am05}.
That
can be overcome by using the, recently developed~\cite{Am03},
multi-channel algebraic scattering (MCAS)
scheme; applications of which, even starting
with interactions built from the simplest, local form, collective model
descriptions, have been quite successful~\cite{Ca05,Am03,Pi05}.
This method of solution of coupled-channel scattering problems 
has a number of salient features, notably:
\begin{enumerate}
\item 
With this method, it is quite straight-forward to define
specific radial forms of the $NA$ optical potentials 
which account for the effects of all channels considered~\cite{Ca91}.
\item 
Pauli exclusion effects can be included even when simple (local) collective
models are used to define the initial potential function matrix $V_{cc'}(r)$.
\item
All sub-threshold (bound) states of the compound system can be defined;
\item
There is a procedure by which the centroids and widths of all resonances 
in the scattering can be determined, no matter how narrow any resonance may be.
\item 
It can be used with any matrix of initial potentials, whether they be local
or non-local in form. The latter will be the case, due to the Pauli principle, 
when detailed nuclear structure is used to define the initial matrix of
interaction potentials.
\item
$S$-matrices and, in fact, complete off-shell $T$-matrices are evaluated; 
the latter 
required for analysis of off-shell processes such as bremsstrahlung or
nucleon capture reaction cross sections.
\end{enumerate}

In the next section, we present some theoretical considerations, outlining
the methods we have used to analyze scattering data and specify
the effects of non-locality in the problems. For the case 
of low-energy scattering, we also identify the non-locality in the optical 
potential. 
Then, in Sec.~III, 
we present and discuss results. Finally, in Sec. IV, conclusions are drawn.


\section{Theory considerations}

An elegant formulation of $NA$
optical potentials,  given in the book by Feshbach~\cite{Fe92}, 
has an operator specification  of
the dynamical polarization potential (DPP).
The Feshbach formalism uses projection operators that
divide the Hilbert space into the channels of a given scattering 
problem that are to be considered explicitly, $\mathcal{P}$, and all other 
channels, $\mathcal{Q}$. 
Feshbach determined that the Schr\"odinger equation for scattering
took the form
\begin{equation}
\left(E - H_{{\mathcal P}{\mathcal P}} -
H_{{\mathcal P}{\mathcal Q}} 
\left[E - H_{{\mathcal Q}{\mathcal Q}} + i\epsilon\right]^{-1} 
H_{{\mathcal Q}{\mathcal P}}
  \right)
\left| \Psi^{(+)} \right>  = 0\ ,
\end{equation}
where
$H_{{\mathcal P}{\mathcal Q}} \equiv {\mathcal P}H{\mathcal Q}$
and similarly for $H_{{\mathcal Q}{\mathcal P}}, H_{{\mathcal P}{\mathcal P}}$,
and $H_{{\mathcal Q}{\mathcal Q}}$.
Thus the DPP, formally, is a projection of $\mathcal Q$
space effects to an effective operator in $\mathcal P$ space, i.e.
\begin{equation}
\Delta U_{{\mathcal P}{\mathcal P}}
= 
H_{{\mathcal P}{\mathcal Q}} 
\left[E - H_{{\mathcal Q}{\mathcal Q}} + i\epsilon\right]^{-1} 
H_{{\mathcal Q}{\mathcal P}}
\ .
\end{equation}

We consider the case that $\mathcal{P}$ projects onto the elastic-scattering 
channel, then 
\begin{equation}
\mathcal{P} = \ket{\Psi_{gs}} \bra{\Psi_{gs}}
\; ;\;
\mathcal{Q} = \mathbf{1} - \ket{\Psi_{gs}} \bra{\Psi_{gs}}
\; ;\;
\mathcal{Q} \ket{\Psi_{gs}} = 0\ , 
\end{equation}
and take as the Hamiltonian,
\begin{equation}
H = H_0 + V + H_A \hspace*{0.5cm} ;\hspace*{0.5cm} 
H_A \ket{\Phi_j} = e_j \ket{\Phi_j}\ .
\label{9}
\end{equation}
$\ket{\Phi_j}$ are the eigenstates of the Hamiltonian for the target nucleus 
and $e_j$ are the eigenvalues. 
Then as only the operator $V$ connects $\mathcal Q$ and $\mathcal P$ spaces,
the DPP is
\begin{equation}
\Delta U_{{\mathcal P}{\mathcal P}}
= 
V_{{\mathcal P}{\mathcal Q}} 
\left[E - H_{{\mathcal Q}{\mathcal Q}} + i\epsilon\right]^{-1} 
V_{{\mathcal Q}{\mathcal P}}
\ .
\end{equation}

For simplicity, temporarily, we ignore antisymmetrization 
between the continuum projectile nucleon and all the $A$ nucleons of the 
target~\cite{Fe92}. Then the ($A+1$) particle states have the form
\begin{eqnarray}
\ket{\Psi_{j}^{+}} = \ket{\chi^{(+)}_j} \ket{\Phi_j}
&=& \ket{\chi^{+}_j(0)} \, \ket{\Phi_j(1, \cdots, A)}\ ,
\nonumber\\
{\mathcal P} \ket{\Psi_{j}^{+}} = 
\ket{\Psi_{gs}^{+}} 
&=& \ket{\chi^{+}_{gs}(0)} \, \ket{\Phi_{gs}(1, \cdots, A)}\ ,
\label{10}
\end{eqnarray}
and on taking the target ground-state expectation, 
\begin{eqnarray}
\biggl( E - H_0 &-& e_{gs} - \me{\Phi_{gs}}{V}{\Phi_{gs}} 
\nonumber\\ 
&-& \me{\Phi_{gs}}{ V G^{(+)}_{\mathcal{QQ}} V}{\Phi_{gs}} \biggr)\,
\ket{\chi^{+}_{gs}} = 0 \, ,
\label{11}
\end{eqnarray}
where
\begin{equation}
G^{(+)}_{\mathcal{QQ}} = 
\left[ E - H_{\mathcal{QQ}} + i \epsilon \right]^{-1}
\ . 
\end{equation}
Thus, a structural form for the $NA$ optical potential is identified by
\begin{eqnarray}
U_{OM}(E) &=&  \me{\Phi_{gs}}{V}{\Phi_{gs}} +
\me{\Phi_{gs}}{ V G^{(+)}_{\mathcal{QQ}} V }{\Phi_{gs}}
\nonumber\\
&=& \me{\Phi_{gs}}{V}{\Phi_{gs}} + \Delta U(E)\ .
\label{OMform}
\end{eqnarray}
Here $\Delta U(E)$ is the DPP which leads to a 
coordinate space optical potential,
\begin{equation}
U_{OM}({\bf r},{\bf r'};E) = V_{loc}({\bf r})\, \delta({\bf r} - {\bf r'})
+ \Delta U({\bf r},{\bf r'};E)\ ,
\label{CoordOM}
\end{equation}
when a local form ($V_{loc}$) for the elemental ground-state interaction 
is assumed.

But allowing for the Pauli principle whereby the emergent nucleon may 
not be that incident on the target makes even the leading term
non-local. 
The Schr\"odinger equations for the relative 
motion wave functions thus have the form (with $\chi^{(+)}({\bf r}) \equiv 
\chi^{(+)}({\bf r},E)$)
\begin{equation}
\left(\frac{\hbar^2}{2{\overline m}}\nabla^2 + E \right)\, \chi^{(+)}({\bf r}) =
\int U_{OM}({\bf r},{\bf r'};E)\, \chi^{(+)}({\bf r'})\, d{\bf r'}\ .
\end{equation}
Here ${\overline m}$ is the reduced mass, and the energy scale is
taken with $\epsilon_{gs} = 0$


\subsection{Low-energy regime and MCAS}

For nucleons of up to 5 MeV scattering from light mass targets particularly, 
cross sections show relatively few (resonance) states which are often widely 
spaced.  Low energies also mean few partial waves are important with the 
scattering (both factors keep the scale of the problem of using MCAS 
manageable). Nonetheless, the applicability of MCAS 
is only limited by the computing power and time available.

The essential input to MCAS is a matrix of potentials that
define a specific $NA$ system. The first operational
requirement then is to find the optimal expansion of those potentials
in separable form. That expansion is made in terms of sturmian
functions~\cite{Am03,Ca91} with the sturmians being
generated from the chosen matrix of potentials themselves. Then, with
separable interactions, the Hilbert-Schmidt expansion
of amplitudes gives the $T$-matrix also in separable form.

The MCAS theory~\cite{Am03} treats coupled-channel scattering in momentum 
space, giving
solutions of Lippmann-Schwinger (LS) integral equations.
For each spin parity $J^\pi$ of a given $NA$ system, one must
consider a set of $\Gamma$ scattering channels, each with a label $c$
$(1 \leq c \leq \Gamma)$ where each $c$ identifies a set of quantum numbers
(all details have been published~\cite{Am03}).  Then, in partial wave form, the 
coupled LS equations define a multichannel $T$-matrix,
\begin{eqnarray}
&& T^{J^{\pi}}_{cc'}( p, q; E ) = V^{J^{\pi}}_{cc'}( p, q )
\nonumber\\
&&\hspace*{0.1cm} 
+\ \mu \left[ \sum^{\rm open}_{c'' = 1} \int^{\infty}_0
  V^{J^{\pi}}_{cc''}( p, x ) \frac{ x^2 }{ k^2_{c''} - x^2 + i\epsilon
  } T^{J^{\pi}}_{c''c'}( x, q; E ) \, dx \right. 
\nonumber \\
&&\hspace*{0.4cm} - \left. \sum^{\rm closed}_{c'' = 1} \int^{\infty}_0
  V^{J^{\pi}}_{cc''}( p, x ) \frac{ x^2 }{ h^2_{c''} + x^2 }
  T^{J^{\pi}}_{c''c'}( x,q; E ) \, dx \right] 
  \label{14}
\end{eqnarray}
where $\mu = {2 \overline{m}}/\hbar^2$.
There are two summations as the open and closed channel components have 
been separated with the wave numbers being
$k_c = \sqrt{\mu(E - \epsilon_c)}$ and $h_c = \sqrt{\mu(\epsilon_c - E)}$
for $E > \epsilon_c$ and $E < \epsilon_c$ respectively. $\epsilon_c$ is the
energy threshold at which channel $c$ is open. Note that henceforth the $J^\pi$
superscript is to be understood. Expansion of $V_{cc'}$ in terms of a finite
number of sturmians yields
\begin{equation}
V_{cc'}(p,q) \sim  \sum^N_{n = 1} \hat{\chi}_{cn}(p)\,
\eta^{-1}_n\, \hat{\chi}_{c'n}(q)\; ,
\label{15}
\end{equation}
where ${\hat \chi}_{cn}(p)$ relate to the chosen sturmian functions in momentum
space and $\eta$ are the associated eigenvalues~\cite{Am03,Ca91}.

The Fourier-Bessel transforms of the form factors are $\chi_{cn}(r)$. They
are also defined in terms of the sturmians ($\Phi_{cm}(r)$) by
\begin{equation}
\chi_{cn}(r) =  \sum^\Gamma_{c'=1} V_{cc'}(r) \Phi_{c'n}(r) 
\end{equation}
if the initial potentials are local in form, and by
\begin{equation}
\chi_{cp}(r) = \sum^\Gamma_{c'=1} \int_0^\infty V_{cc'}(r, r') \,
\Phi_{c'p}(r')\, dr' 
\end{equation}
if those potentials  are  non-local.
It is possible to develop the optical potential in coordinate space with the 
MCAS theory~\cite{Am03,Ca91}.  We will consider only spin zero targets 
so that the channel indices designate the collected
angular momenta $c \equiv \left[(l \frac{1}{2})j I;J \right]$. Then,
for each conserved total angular momentum $J$ and selected parity,
the partial wave $l$ associated with a zero spin ground state is
unique. In the following, the index  $c=1$ then implies the 
partial wave specific to each $J^\pi$ considered. 

Assuming a local form for the leading term,
MCAS theory gives for the DPP, 
\begin{equation}
\Delta U_{1, 1}(r, r'; E) 
= \sum^\Gamma_{c,c'=2} V_{1 c}(r)\, G^{(Q)}_{cc'}(r,r';E)\, V_{c' 1}(r')\ . 
\end{equation}
The Green's functions are solutions of
\begin{equation}
G^{(Q)}_{cc'} = G^{(0)}_{c} \delta_{cc'} + 
\sum^\Gamma_{c'' = 2}\ G^{(0)}_{c} \, V_{cc''}\, G^{(Q)}_{c''c'}\ ,
\label{25}
\end{equation}
where $G^{(0)}_{c}$ is the free Green's function for each individual
channel $c$.  Historically, finding these solutions has been difficult. 
However the MCAS development allows the analogous definition~\cite{Ca90,Ca91a},
\begin{equation}
\Delta U_{11}(r,r';E) = \sum^N_{n, n' = 1}  \chi_{1 n}(r) 
\left[{\mbox{\boldmath $\Lambda$}}(E) \right]_{nn'} 
\chi_{1n'}(r')\ ,
\label{26}
\end{equation}
where
\begin{eqnarray}
&&{\mbox{\boldmath $\Lambda$}}(E) = 
\left[ {\mbox{\boldmath $\eta$}} - \mathbf{G}_0^{(Q)}(E)\right]^{-1} 
- {\mbox{\boldmath $\eta$}}^{-1}\ ,
\nonumber\\
&&\left[ \mathbf{G}_0^{(Q)}(E)\right]_{n n'} = \mu \left[
\sum^{\rm open}_{c\ne 1} \int_0^\infty \frac{\hat{\chi}_{cn}(x)
\hat{\chi}_{cn'}(x)}{k_c^2 - x^2 + i\epsilon}\, x^2\, dx \right.
\nonumber\\
&&\left.\hspace*{2.1cm}
- \sum^{\rm closed}_{c\ne 1} \int_0^\infty \frac{\hat{\chi}_{cn}(x)
\hat{\chi}_{cn'}(x)}{h_c^2 + x^2}\, x^2\, dx \right] .
\label{27}
\end{eqnarray}

\subsubsection{The Perey-Buck (${\rm PB}$) non-local potential}

The PB potential~\cite{Pe62} has the energy-independent, non-local, form
\begin{eqnarray}
U_{OM}^{PB}(\textbf{r},\textbf{r}';E) &=& 
v\left( {\textstyle\frac{1}{2}}|\textbf{r}+\textbf{r}'| \right) 
\frac{1}{[\sqrt{\pi} 
\beta_{NL}]^3} e^{-\left( \frac{\textbf{r}-\textbf{r}'}{\beta_{NL}} \right)^2}\
\nonumber\\
&&\Rightarrow\ v(s)\frac{1}{[\sqrt{\pi} \beta_{NL}]^3}
e^{ - \left( \frac{\textbf{r}-\textbf{r}'}{\beta_{NL}} \right)^2}\ ,
\label{VPB}
\end{eqnarray}
where, with $s = {\textstyle\frac{1}{2}}(\textbf{r}+\textbf{r}')$, 
\begin{equation}
v(s) = 
V_{NL} \left[ 1 + e^{ \left( \frac{s - R_{NL}}{a_{NL}} \right)} \right]^{-1}.
\end{equation}
As shown in the appendix of Ref.~\cite{Pe62}, 
this (reduced) form can be 
expanded in partial waves with radial multipoles being
\begin{eqnarray}
\frac{1}{rr'}g_{\ell}(r,r') &=& 
\frac{2}{\sqrt{\pi} \beta_{NL}^3}\ 
v \left( {\textstyle\frac{1}{2}}[r + r'] \right)
\nonumber\\
&&\hspace*{0.5cm}\times
\ e^{- \left[ \frac{(r^2 - r'^2)}{\beta^2_{NL}} \right]}
i^{\ell}j_{\ell} \left( -i \frac{2rr'}{\beta^2_{NL}} \right)\ .
\end{eqnarray}
However, there are parameters involved and so this potential is 
phenomenological.  Nonetheless, the parameters of this interaction, intended 
for use at low energies and with nuclei of medium and heavy mass, were 
obtained solely by fitting the differential cross sections for 7.0 and 14.5 
MeV neutrons elastically scattered from ${}^{208}$Pb. Those parameter values 
then were used~\cite{Pe62} in analyses of a set of neutron scattering 
observables for a range of energy, 0.4 to 24.0 MeV, and for a set of targets 
ranging from ${}^{27}$Al to ${}^{208}$Pb. Good agreement was found between 
results obtained with those parameters and data taken with those targets. 

Though the PB interaction was not used for scattering from ${}^{12}$C,
we adopt it in this study to display its non-local features.
The PB prescription is energy 
independent and so the shape of that non-locality is constant. 
The results may not be the  PB potential that gives
a best fit to scattering data from ${}^{12}$C. But as we only seek
to make qualitative comparison with the MCAS results  it suffices to use the parameter values initially defined~\cite{Pe62} .

\subsection{The $g$-folding model for medium energies}
\label{dwa-theory}

For energies well above the particle-emission threshold and so
coinciding with a continuum of states in the target, multi-nucleon 
scattering reductions of the formal Feshbach theory of scattering
are relevant.  It is useful to follow the KMT scheme~\cite{Ke59} 
in which it is assumed that only 
pairwise interactions between the projectile and individual target nucleons 
are 
important.  With the projectile tagged by the subscript $0$, and 
target nucleons tagged similarly by
$i$, the essential interaction can be written 
\begin{equation}
V = \sum_{i = 1}^A v_{0i} = A v_{01}\; .
\end{equation}
With whatever is chosen for the free $NN$ interaction $v_{01}$, 
the KMT theory allows for multiple pairwise interactions so that
$NN$ $t$-matrices, solutions of 
$NN$ LS equations, are required in applications. However, such an approach 
does not 
account for the
ways in which those interactions are influenced by the nuclear medium
in which the two nucleons interact. Experience~\cite{Am00} has shown that 
the $NN$ 
$g$-matrices, solutions of the Brueckner-Bethe-Goldstone (BBG) equations 
for infinite nuclear matter~\cite{Be63}, can well approximate those 
(many-body) corrections.
Using the $NN$ $g$-matrices in a first order KMT approach involves an
assumption that the important terms in Q-space, for the incident energies
concerned, are excitations in which a 
particle is promoted to a state in an infinite-matter system with
Fermi momentum related to the density of the nucleus from where it came. 
Thus, the effective $NN$ interaction, besides being energy dependent
and complex, also is density dependent~\cite{Am00}.  
The optical potentials that result on folding these interactions with
density matrices of the target nucleus, will be both complex and 
energy-dependent.  Further, by
virtue of the Pauli principle at least, they are non-local.

The $g$-matrices can be used directly within a momentum
space formulation of the $NA$ optical potential~\cite{Ar96} or by forming an 
effective $NA$ interaction in coordinate space~\cite{Am00}.
Our approach is with the latter and requires
a mapping of the actual $NN$ $g$-matrices
into the coordinate space forms usable in the program DWBA98~\cite{Ra98}.

In coordinate space, such non-local optical potentials can be written formally
\begin{equation}
U_{OM} = U_{OM}(\vec{r}_1,\vec{r}_2; E)
=  U_{OM}^D(\vec{r}_1; E) + U_{OM}^{Ex}(\vec{r}_1,\vec{r}_2; E)
\end{equation}
where
\begin{eqnarray}
U_{OM}^D(\vec{r}_1; E)
&=& \sum_n \zeta_n \,  \delta(\vec{r}_1 - \vec{r}_2) 
\int \varphi^{\ast}_n(\vec{s}) \, v_D( \vec{r}_{1s} ) \,
\varphi_n(\vec{s}) \; d^3s
\nonumber\\
&=&  \delta(\vec{r}_1 - \vec{r}_2) 
\int \rho({\vec s})\, 
 v_D( \vec{r}_{1s} ) \, d^3s
\end{eqnarray}
involving then an integration over the nuclear density, and the exchange
term which is totally nonlocal,
\begin{equation}
U_{OM}^{Ex}(\vec{r}_1,\vec{r}_2; E)
= \sum_n \zeta_n \, \varphi^{\ast}_n(\vec{r}_1) \,
v_{Ex}(\vec{r}_{12}) \, \varphi_n(\vec{r}_2)\ .
\label{NonSE}
\end{equation}
$v_D$ , $v_{Ex}$ are combinations of the components of the
effective $NN$ interactions, $\varphi_n(\vec{r})$ are single-nucleon 
bound states, and $\zeta_n$ are bound-state shell occupancies. More
generally, the latter  are one-body density matrix elements (OBDME).
It is not easy to specify actual radial values of these non-local
potentials for graphing purposes.  The program
DWBA98~\cite{Ra98} does not form such in evaluation of solutions
of the partial wave Schr\"odinger equations.
The operator structure and strong density 
dependence of the effective interactions deduced from the $NN$ $g$-matrices 
do not lend themselves easily to create multipole
expansions needed to specify $U_{OM}(r_1,r_2)$ for each partial wave. 
In finding solutions of the integro-differential equations, the program 
DWBA98 uses expansions involving particle-hole expectation values of the 
effective interactions~\cite{Am00}.  However, the effects of the  
inherent non-locality
are clearly evident from calculated cross sections and spin observables
as we show later. They are also important in predictions of cross-section
and spin-observable data from inelastic scattering of nucleons
from nuclei.

Cross sections for inelastic proton scattering have  been  evaluated
using  a microscopic DWA theory of the processes~\cite{Am00}.  
In that theory, the transition amplitudes for 
nucleon inelastic scattering  from a nuclear target 
have the form~\cite{Am00} ($\Psi_{J_iM_i} \equiv \Psi_{J_iM_i}(1\cdots A)$)
\begin{eqnarray}
{\cal T} &=& T^{M_fM_i\nu^\prime\nu}_{J_fJ_i}(\Omega_{sc})
\nonumber\\
&=& \left\langle \chi^{(-)}_{\nu^\prime}({\bf k}_o 0)\right|
\left\langle\Psi_{J_fM_f} \right|
\; A\ {\bf g}_{\rm eff}(0,1)
\nonumber\\
&&\hspace*{2.2cm} \times
{\cal A}_{01} \left\{ \left| \chi^{(+)}_\nu ({\bf
k}_i0) \right\rangle \right.  \left. \left| \Psi_{J_iM_i}
\right\rangle \right\} ,
\end{eqnarray}
where $\Omega_{sc}$ is the scattering angle and  ${\cal A}_{01}$ is 
the two-nucleon state antisymmetrization operator. 
The nuclear transition is from a state $\vert J_i M_i \rangle$
to a state $\vert J_f M_f \rangle$ and the projectile has spin projections
$\nu$ before, and $\nu^\prime$ after, the collision. The incoming
and outgoing distorted waves are specified by $\chi^{\pm}$ and they have
relative momenta of ${\bf k}_i$ and ${\bf k}_o$ respectively.  The 
development proceeds by using a 
cofactor expansion of the target states, 
\begin{equation}
\left| \Psi_{JM} \right\rangle  
= \frac{1}{\sqrt{A}} \sum_{j,m}
\left| \varphi_{jm}(1) \right\rangle\,
a_{jm}(1)\, \left| \Psi_{JM}\right\rangle\ ,
\label{cofactor}
\end{equation}
which allows  expansion of the  scattering  amplitudes in the form  of 
weighted two-nucleon elements since 
$a_{jm}(1)\, \left| \Psi_{JM}\right\rangle$ in 
Eq.~(\ref{cofactor}) is independent of coordinate `1'. Thus 
\begin{eqnarray}
{\cal T} &=& \sum_{j_1,j_2} \left\langle\Psi_{J_fM_f} \right|
a^{\dagger}_{j_2m_2}(1)\ a_{j_1m_1}(1) \left| \Psi_{J_iM_i} 
\right\rangle 
\nonumber\\
&&\; \times
\left\langle \chi^{(-)}_{\nu^\prime}({\bf k}_o0)\right|
\left\langle \varphi_{j_2m_2}(1) \right| \ {\bf g}_{\rm eff}(0,1) 
\nonumber\\
&&\hspace*{2.0cm}
{\cal A}_{01} \left\{ \left| \chi^{(+)}_\nu ({\bf k}_i0) 
\right\rangle
\ \left| \varphi_{j_1m_1} (1) \right\rangle \right\}
\nonumber\\
&=& \sum_{j_1,j_2,I}
\frac{1}{\sqrt{2J_f+1}}\,  
\left< J_i\, I\, M_i\, N \vert J_f\, M_f \right>
\; S_{j_1\, j_2\, I}^{(J_i \to J_f)}
\nonumber\\
&&\ \times 
\sum_{m_1,m_2}
(-1)^{(j_1-m_1)} 
\left< j_1\, j_2\, m_1\, -m_2 \vert I\, -N \right>
\nonumber\\
&&\;\times 
\Bigl< \chi^{(-)}_{\nu^\prime}({\bf k}_o0)\Bigr|
\left\langle \varphi_{j_2m_2}(1) \right| \
{\bf g}_{\rm eff}(0,1)
\nonumber\\
&&\hspace*{2.0cm}
 {\cal A}_{01} \left\{ \left| \chi^{(+)}_\nu ({\bf k}_i0) 
\right\rangle
\ \left| \varphi_{j_1m_1} (1) \right\rangle \right\} ,
\end{eqnarray}
where reduction of the structure factor to (transition) OBDME 
$\left( S_{j_1\, j_2\, I}^{(J_i \to J_f)} \right)$ 
for  angular momentum transfer values $I$ is given in detail  
elsewhere~\cite{Am00}.

The effective interactions $g_{\rm eff}(0,1)$ used in the folding to
get  the  optical  potentials   have   also   been   used  as  the  
transition  operators  effecting  the excitations. 
As with the generation of the elastic-scattering optical
potentials from which the distorted waves are generated,
antisymmetry   of   the  projectile  with  each individual  bound 
nucleon is treated exactly.   The associated knock-out (exchange) 
amplitudes are the non-local effects in inelastic scattering. They 
 contribute   importantly   to  the  evaluated scattering  cross  
section, both in magnitude and shape~\cite{Am00}.

\section{Results and Discussion}

\subsection{MCAS and the low-energy regime}

In the paper detailing the MCAS method~\cite{Am03} 
cross sections and spectra from the scattering of neutrons from
$^{12}$C for energies to $\sim 6$ MeV were studied.
The coupled-channel starting potentials were defined from a 
collective model which included quadrupole deformation.  The spectrum 
of the target, $^{12}$C, was truncated to just the lowest three 
states; the $0_1^+$ ground state, the $2^+$ state
at 4.43 MeV and the $0_2^+$ state at 7.96 MeV.  We consider those results 
again in brief as they are the bound and scattering properties that the 
DPP we form will reproduce when used in a non-local Schr\"odinger equation.

\begin{figure}
\begin{center}
\scalebox{0.4}{\includegraphics*{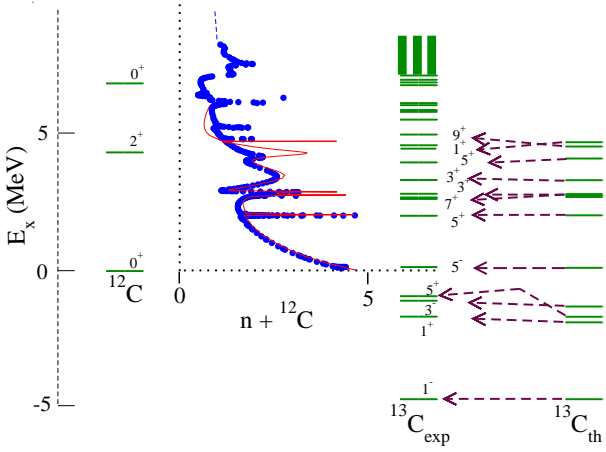}}
\end{center}
\caption{\label{Fig1} (Color online)
MCAS results compared with data  from the elastic $n$+${}^{12}$C scattering
and with the spectrum of ${}^{13}$C. Each ${}^{13}$C state 
is identified by their parity and twice their spin.}
\end{figure}

With parameter values as used previously~\cite{Am03}, MCAS calculations gave 
the 
energy variation cross section for low energy $n$+${}^{12}$C elastic
scattering that is shown by the solid curve in Fig.~\ref{Fig1}. 
The filled circles are the data~\cite{Pe93,IA05}. The compound nucleus
sub-threshold (bound) and resonance states that result
are compared with the ${}^{13}$C spectrum therein as well.
The comparison of calculated results with data is very good at least to the 
energy (4.43 MeV) coinciding with excitation of the first $2^+$ state. 
It is of note that, using MCAS for the negative (sub-threshold) energy 
regime gives the correct bound states of ${}^{13}$C. Accounting for the
Pauli principle was crucial to find these results; just how crucial
was shown in recent publications~\cite{Ca05,Am05}. Adding a Coulomb
interaction also gave a good spectrum for ${}^{13}$N and good reproduction
of fixed angle, proton elastic-scattering observables~\cite{Pi05}.

\begin{figure}
\begin{center}
\scalebox{0.45}{\includegraphics*{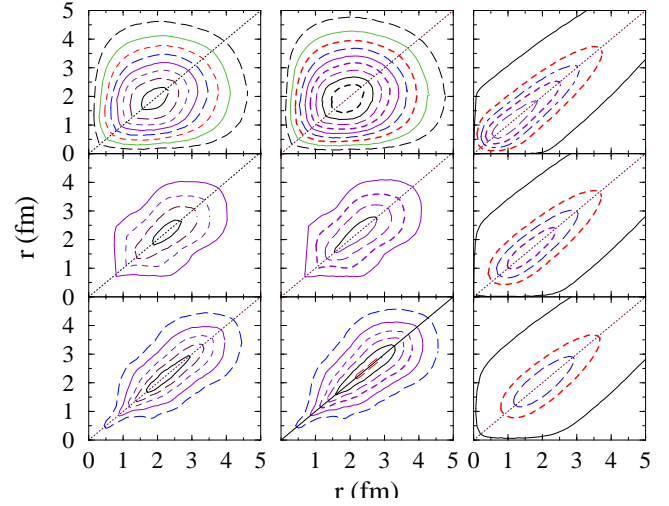}}
\end{center}
\caption{\label{Fig2} (Color online)
Contours of the DPP found using MCAS for neutrons of energies 1.5 MeV 
(left) and 2.73 MeV (middle) compared with a PB energy-independent
form (right).}
\end{figure}

We have evaluated the DPP for this $n$+${}^{12}$C system and results are
shown in Fig.~\ref{Fig2}. Therein the $0s_{\frac{1}{2}}$, 
$0p_{\frac{3}{2}}$, and $d_{\frac{5}{2}}$ wave DPP values
are plotted  in the top, middle and bottom segments for energies 
of 1.5 MeV  and of 2.73 MeV 
in the left and middle panels respectively. The right panel
contains some PB potential results that we discuss later.
These DPP display a well shape with maximal depths
on axis and in the nuclear surface region. 
The units for these DPP are MeV-fm$^{-1}$ and the 
$0s_{\frac{1}{2}}$ wave
contours have spacings of  20 MeV-fm$^{-1}$ from a value of $-20$ MeV-fm$^{-1}$ 
with the most outer (dashed) curves. The central well depth of these
$0s_{\frac{1}{2}}$ wave DPP are $-170$ MeV-fm$^{-1}$ and $-240$ MeV-fm$^{-1}$ for 
1.5 MeV and 2.73 MeV neutrons respectively.  The $0p_{\frac{3}{2}}$ and 
$0d_{\frac{5}{2}}$ wave DPP respectively are given in contours
starting with the outermost value of $-5$ MeV-fm$^{-1}$ with $-5$ MeV-fm$^{-1}$ 
steps inward
to maximal depth values of $-22 (-23)$ MeV-fm$^{-1}$ and $-28 (-32)$ MeV-fm$^{-1}$ 
for
the incident energy of 1.5 (2.73) MeV.
Clearly the DPP are strongly non-local and energy dependent;
the $0s_{\frac{1}{2}}$ wave for this case of $n$-${}^{12}$C scattering 
particularly so.
The $0p_{\frac{3}{2}}$ and $0d_{\frac{5}{2}}$ wave forms are not as energy 
dependent over the 0 to 4 MeV 
projectile energy range but they, too, are markedly non-local.

\begin{figure}
\begin{center}
\scalebox{0.41}{\includegraphics*{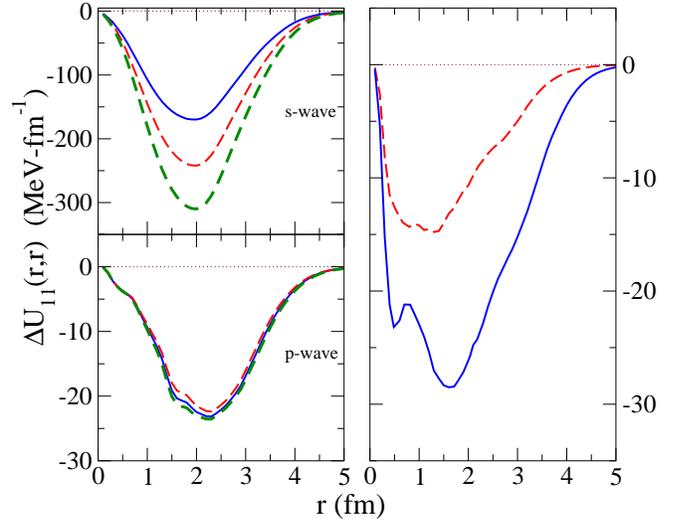}}
\end{center}
\caption{\label{Fig3} (Color online)
The radial variations of the DPP along the diagonal for 
$0s_{\frac{1}{2}}$ waves (top left), 
$0p_{\frac{3}{2}}$ waves (bottom left) and for $0d_j$ waves (right).
Details are given in the text.}
\end{figure}

Values of DPPs along the diagonal ($r = r'$) are shown in Fig.~\ref{Fig3}. 
The $0s_{\frac{1}{2}}$ and $0p_{\frac{3}{2}}$ wave DPP diagonal values, shown 
in the top and in the bottom (left) panels respectively, are for incident 
energies of 1.5 
MeV (solid), 2.73 MeV (long dashed), and 3.23 MeV (dashed curves). 
These clearly show the strong energy dependence of the $0s_{\frac{1}{2}}$ wave 
interactions and the almost no energy variation in the 
$0p_{\frac{3}{2}}$ wave DPP.
The well shapes do not change with energy and the maximal depths are located
at 2.0 and 2.2 fm respectively for the $0s_{\frac{1}{2}}$ and 
$0p_{\frac{3}{2}}$ wave results. 
In the right panel of Fig.~\ref{Fig3}, $0d_j$ wave DPPs for the energy of  
2.73 MeV are displayed.
The results for $j = \frac{3}{2}$ and $\frac{5}{2}$ are depicted by the
dashed and solid curves respectively, illustrating  
the spin-orbit attributes formed using the MCAS approach.   
These variations were found~\cite{Am03} without considering spin dependent 
scattering data in the determination of the matrix of initial potentials  
that are input to the MCAS approach. 
Rather it was the spectrum (bound and resonant) of the compound nucleus
(${}^{13}$C) that set the parameterization.
Nonetheless spin dependent scattering  data were well predicted
by this model~\cite{Am03,Sv06}.

\subsubsection{Results using the PB non-local potential}

As the non-locality of this model is energy independent, the only question 
is of what parameter values to use. 
As there are no parameter values available that best fit relevant data,
for simplicity we have chosen those
used by Perey and Buck in their calculations with heavier nuclei.
As noted, those parameter values suffice since 
as we seek a qualitative, but not quantitative,  comparison of the features 
of the PB potentials with those found from MCAS. 
By using  
$V_{NL} = -74$ MeV, $R_{NL} = 2.8$ fm, $a_{NL} = 0.65$
fm and $\beta_{NL} = 0.85$~fm, calculations of Eq.~(\ref{VPB}) gave contour 
plots shown in the right panel of Fig.~\ref{Fig2}. 
The contour lines depict energy spacings of 10~MeV-fm$^{-1}$ 
with the outermost contour that for $-0.1$~MeV-fm$^{-1}$.  The central depths of these 
partial wave interactions are $-46.1$, $-36.4$, and $-28.7$ MeV-fm$^{-1}$ for the 
$0s_{\frac{1}{2}}$, $0p_{\frac{3}{2}}$, and $0d_{\frac{5}{2}}$ waves.
Similarities exist between these non-local potentials and those found
from low-energy MCAS calculations of the DPP. 
The depths of the wells for the two studies are comparable, with 
the depths of the DPP at 2.73 MeV being close to the depths of the 
PB potential for the $0p_{\frac{3}{2}}$ and $0d_{\frac{5}{2}}$ waves.  
The $0s_{\frac{1}{2}}$ wave MCAS depth is
3 times larger than the PB value however. 
Nonetheless, it seems that the PB model does have non-locality
qualitatively similar to what MCAS yields but the energy
independence of the PB model's parameters seems too restrictive.

\subsubsection{Energy above the first excited state}

\begin{figure}[h]
\begin{center}
\scalebox{0.45}{\includegraphics*{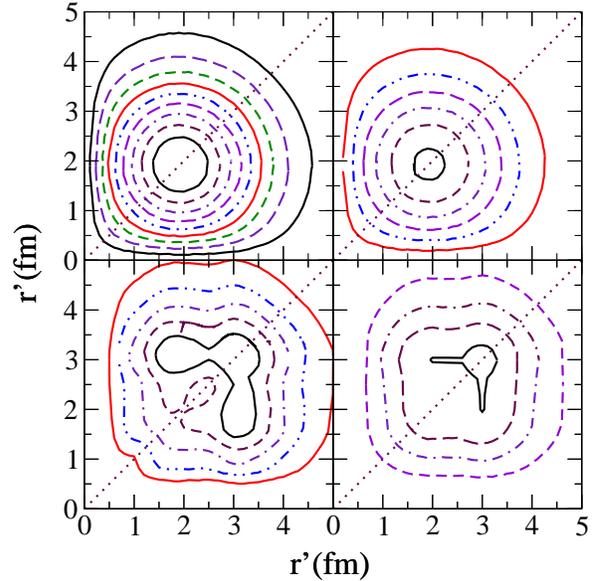}}
\end{center}
\caption{\label{Fig4} (Color online) Contour plots of the DPP
for $0s_{\frac{1}{2}}$ wave (top) and 
$0p_{\frac{1}{2}}$ wave (bottom) at 5.0 MeV.
The real and imaginary parts are shown on the left and right 
of each.}
\end{figure}

For energies where more than the elastic channel is open (above 4.43 MeV
in the case of ${}^{12}$C), there is the possibility of flux loss to
inelastic scattering. Therefore the DPP become complex. At 5.0
MeV in the $n$+${}^{12}$C system, the MCAS method gives the complex
DPP for the $0s_{\frac{1}{2}}$ and $0p_{\frac{1}{2}}$ waves that are 
displayed on the top and bottom in Fig.~\ref{Fig4} respectively.
Contours of the real and imaginary terms are shown on
the left and right respectively in each case.  
The real parts of these potentials at this energy are repulsive with
the central strengths of 5000 and 58.8 MeV-fm$^{-1}$ for the $s$  and
$p$ wave cases respectively. The imaginary parts of both potentials 
are wells with minima of $-620$ and $-8.6$ MeV-fm$^{-1}$ respectively. 
The contours for the real parts of these potentials are shown for every 
 500 (10) MeV-fm$^{-1}$ with the  $0s_{\frac{1}{2}}$ 
($0p_{\frac{3}{2}}$) wave plots while the imaginary potentials
for each indicate changes of 100 (2) MeV-fm$^{-1}$.
Clearly the forms of these potentials, as well as their being complex, 
have changed markedly from those found for nucleons with energies below 
that of the threshold of the first excited state in ${}^{12}$C.
There is also distinctively new structure in the 
$0p_{\frac{1}{2}}$ DPP.

\begin{figure}[h] 
\begin{center}
\scalebox{0.45}{\includegraphics*{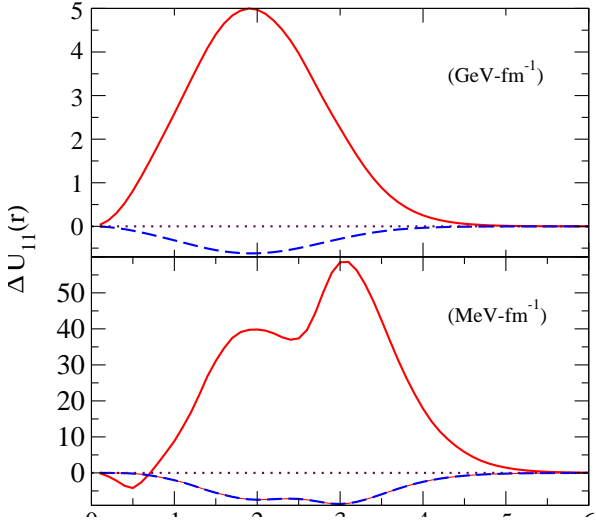}}
\end{center}
\caption{\label{Fig5} (Color online) The diagonal values of the DPPs
obtained using MCAS for 5.0 MeV neutrons scattering from ${}^{12}$C. 
The $0s_{\frac{1}{2}}$  and $0p_{\frac{1}{2}}$ wave potentials are
depicted in the top and bottom sections respectively.
Details are given in the text.}
\end{figure}

The new structures of the DPP at 5.0 MeV are emphasized with the plots
of the diagonal values that are shown in Fig.~\ref{Fig5}.
In the top panel, we show the $0s_{\frac{1}{2}}$ wave potential along the
diagonal. The real part is depicted by the solid curve, the imaginary part 
by the dashed curve.
The bottom section of the diagram contains the real and imaginary
components of the $0p_{\frac{1}{2}}$ wave DPP along the diagonal. Again
the real (imaginary) parts are given by the solid (dashed) curves
therein.  Note that the potential scales are shown in the brackets in
each panel.  Both DPP have large repulsive real parts with the 
$0s_{\frac{1}{2}}$ wave potential being particularly strong. Both also
have absorptive imaginary parts. The especial structure of the 
$0p_{\frac{1}{2}}$ case, already noted in Fig.~\ref{Fig4}, is most
evident in this plot. 

\begin{figure}[h]
\begin{center}
\scalebox{0.4}{\includegraphics*{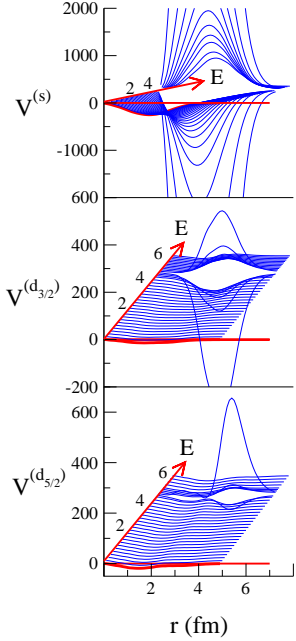}}
\end{center}
\caption{\label{Fig6} (Color online) Energy variation of the 
DPP diagonal potentials for the $s_{\frac{1}{2}}$ wave (top), for the 
$d_{\frac{3}{2}}$ wave (middle), and for the $d_{\frac{5}{2}}$ 
wave (bottom).}
\end{figure}

\begin{figure}[h]
\begin{center}
\scalebox{0.4}{\includegraphics*{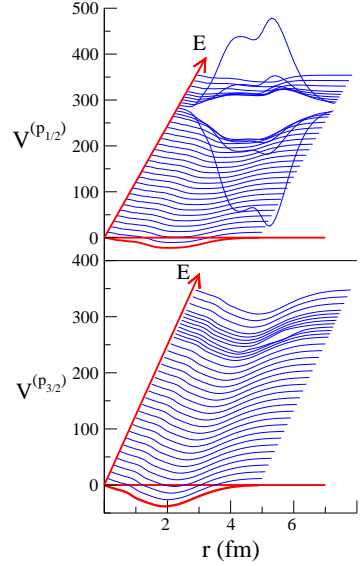}}
\end{center}
\caption{\label{Fig7} (Color online) Energy variation of the
DPP diagonal potentials for the $p_{\frac{1}{2}}$ wave (top) and for  the
$p_{\frac{3}{2}}$ wave (bottom).}
\end{figure}

Finally in Figs.~\ref{Fig6} and \ref{Fig7}, we show the energy
variations of the diagonal ($r = r'$) real parts of the DPPs. Those for
the $s_{\frac{1}{2}}$-, $d_{\frac{3}{2}}$-, and $d_{\frac{5}{2}}$-waves 
are depicted in the top, middle, and bottom of Fig.~\ref{Fig6} respectively.
Note that all of these strengths are given in units of MeV-fm$^{-1}$.
Below the threshold (4.43 MeV) the potentials are purely real and
attractive. 
There is quite strong energy variation in these DPP. At low energies,
the DPP have the form of potential wells with minima in the nuclear
surface region and those well
depths increase with energy, most markedly as the energy approaches
a value of about 4.8 MeV. For higher energies, those real potentials are
strongly repulsive with strength gradually decreasing. 
Passing through the threshold energy, the DPP acts like a potential
wall at or about the nuclear surface. From the specification in 
Eqs.~(\ref{26}) and (\ref{27}), clearly the sharp change in the
character of the real parts of these DPP do not necessarily occur at the
threshold energy. Depending upon the sturmians and their eigenvalues
the scale of change as well as the energy onset can vary with 
spin-parity of the scattering channel. With the latter, there seems to
be a shift to an energy near 4.8 MeV for each case. However the size of
change is very spin-parity dependent.

The diagonal values of the real parts of the DPP for 
the  $p_{\frac{1}{2}}$  and $p_{\frac{3}{2}}$ waves are depicted on top and
bottom of Fig.~\ref{Fig7} respectively. 
In both, the variation with energy across threshold in the
$p_{\frac{1}{2}}$ case is similar to that for the $d_{\frac{3}{2}}$
wave though a double well aspect is evident now.
The threshold effects upon the $p_{\frac{3}{2}}$ wave DPP
show changes but they are
much less dramatic than in the DPP for the other partial waves.

\subsection{Non-locality effects for higher energy data}

The cross section and analyzing power for nucleons elastically, and
inelastically (to the $2^+$, 4.43 MeV state), scattered from $^{12}$C have 
been calculated using OBDME obtained from a complete $(0+2)\hbar\omega$ space 
shell-model calculation~\cite{Ka95}.  Both WS and harmonic oscillator (HO) 
functions have 
been used with those (no-core shell-model) OBDME to calculate the ${}^{12}$C 
optical potentials.  The transition operators for both elastic and inelastic 
scattering processes are the medium dependent effective interactions at the 
relevant energies as described in Sect.~\ref{dwa-theory}. 
All of those elements are used in the 
DWBA98 program~\cite{Ra98}.  This distorted wave approximation (DWA) code
includes exact evaluation of two-nucleon exchange amplitudes that define the 
knock-out process; amplitudes that are the result of non-local aspects of the
reactions since they involve the full nucleon density matrices of 
the target.

With all details preset, just one calculation is made of both the elastic and 
inelastic observables for which the process then gives predictions. However, 
the elements themselves need assessment of their quality. Much 
use~\cite{Am00} has established appropriate effective $NN$ 
interactions in the nuclear medium
as well as of the DWA method (when exchange amplitudes are treated exactly).
Thus one need only assess, by other means if possible, the quality of the
assumed structure. One excellent way to do that is to use the structure in
analyses of electron scattering form factors. 

\subsubsection{An appropriate model of structure}

\begin{figure}[h]
\begin{center}
\scalebox{0.41}{\includegraphics*{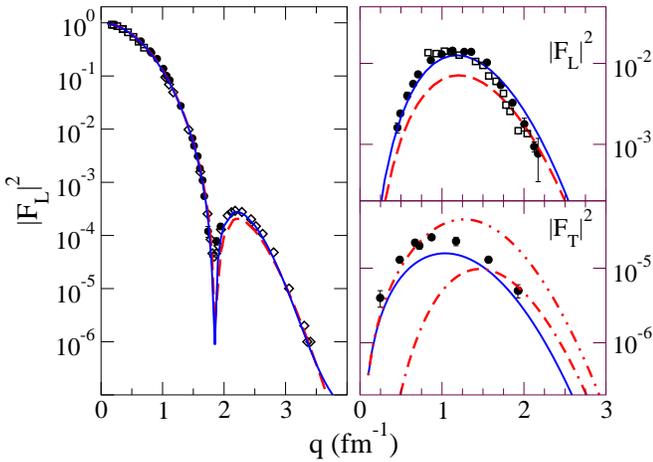}}
\end{center}
\caption{\label{Fig8} (Color online)
Electron scattering form factors for ${}^{12}$C. The elastic longitudinal
form factor is shown in the left panel while the inelastic ones (from
excitation of the $2^+$ (4.43 MeV) state are shown on the right; 
longitudinal (top) and transverse electric (bottom).
All details are given in the text.}
\end{figure}

For the specific case we study,
${}^{12}$C, the longitudinal form factors for elastic and inelastic 
($0^+ \to 2^+$) reactions, and the transverse electric form factor with the 
latter, have been measured accurately over a reasonable range of momentum 
transfer values. Those form factors are compared with our calculated ones
in Fig.~\ref{Fig8}.
Considering the elastic-scattering results (left panel) first, the data 
are those of Jansen {\it et al.}~\cite{Ja72} (squares), of Sick 
and McCarthy~\cite{Si70} (diamonds), and of Nakada {\it et al.}~\cite{Na71}
(circles).  The dashed and solid curves are the results found using WS single
particle wave functions with shell occupancies from 
$0\hbar \omega$ and $(0+2)\hbar\omega$ shell-model calculations
respectively.
The WS potential parameters used to specify the bound-state wave functions
and the shell occupancies are given in the review~\cite{Am00}. 
Similar results were found using HO single-nucleon wave functions when the
oscillator length was 1.6 fm.   
All results  agree well with the data 
though the higher momentum transfer values favor the larger space model
of structure.  But the form factors for the excitation of the first excited
($2^+$) state given in the right panel of Fig.~\ref{Fig8}
differentiates more strongly. 

In the top part of this panel, we present data and results for the
longitudinal electric form factor.
The data displayed by the open squares and filled circles are two sets
reported by Flanz {\it et al.}~\cite{Fl78}.  They  are compared with 
results found using shell-model transition OBDME from the $0\hbar\omega$ 
(dashed curve) and from the $(0+2)\hbar\omega$ (solid curve) shell-model 
calculations.  Clearly the additional contributions from transitions out 
of the $0p$ shell that result in the larger space shell-model study give 
the extra transition strength required to match observation.  

The solid curve shown in the bottom panel is the transverse electric
form factor calculated 
using the $(0+2)\hbar\omega$ shell-model transition OBDME and    WS wave 
functions.     The match to data~\cite{Fl78} is good especially when one 
notes that the separate proton and neutron contributions to this    form 
factor (shown by the dash-double dotted and dot-double dashed curves 
respectively)    have 
amplitudes  that  interfere destructively to determine the total result. 
In the definition of the transition form factor, Siegert operators  were 
used~\cite{Am00} to account for meson exchange current corrections.  
Thus we have confidence that the  no-core $(0+2)\hbar\omega$ 
shell model describes well the ground state of ${}^{12}$C and the 
excitation of its first excited state, as well as of using those wave functions 
and OBDME in analyses of nucleon scattering from ${}^{12}$C. We consider 
two cases; those for 95 MeV neutrons and for 200 MeV protons.

\subsubsection{Credibility of the effective interaction}

\begin{figure}[h]
\begin{center}
\scalebox{0.35}{\includegraphics*{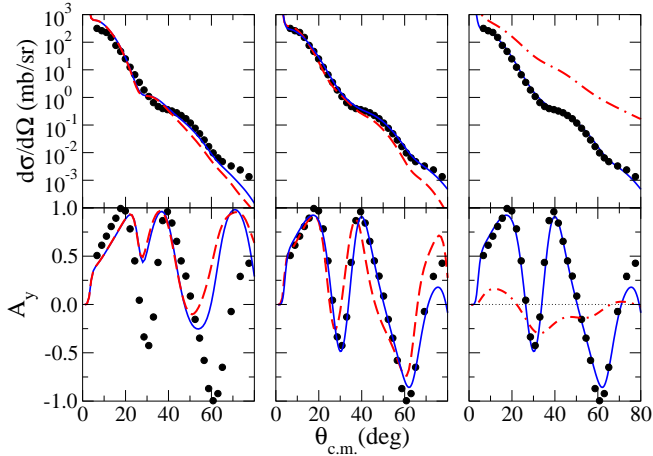}}
\end{center}
\caption[]{\label{Fig9} (Color online)
200 MeV proton elastic-scattering cross section (top) and
analyzing power (bottom). Data~\cite{Co82} (circles) are compared 
with the results found using
the $t$- (left) and $g$-folding (middle) optical potentials
The right panel shows the effect of
omitting the knock-out amplitudes of the $g$-folding potentials.
All details are given in the text.}
\end{figure}

The differential cross sections and analyzing powers for elastic scattering
of 200 MeV protons from ${}^{12}$C that result on using 
the full non-local optical potentials are compared with data~\cite{Co82} 
in Fig.~\ref{Fig9}.  Consistent with 
the results found with the elastic electron scattering form factors, using 
the Cohen and Kurath~\cite{Co65} $0p$ shell-model structure makes but little 
change to the proton elastic-scattering calculations from those displayed.
In the left and middle panels in this figure,
the solid and long dashed curves display the 
results obtained by using WS and HO bound-state wave functions.  Those 
panels contain results found from
calculations made using $t$- (left) and $g$- (middle) folding.
In $t$-folding, the purely free two-nucleon
$t$-matrices were used as the effective interaction. Thus comparing the
results of these two panels shows how important are the
medium modifications of the effective $NN$ interaction. 
It is evident that $g$-folding  with WS bound states gives an optical 
potential from which a differential cross section is obtained that matches 
the data best.  The importance of medium modification in the effective 
$NN$ interaction  is even more obvious with the analyzing power. The 
$g$-folding results are in much better agreement with data than are the 
$t$-folding ones.  However, with this observable as well,  there is 
little to choose between the results obtained using HO and WS bound states
for the single-particle bound states of ${}^{12}$C.

The effect of omitting the exchange amplitudes in defining the elastic
scattering are shown in the rightmost panel. Therein the solid curves
are the complete $g$-folding model results also shown in the middle
panel, while the dot-dashed curves are the results when the exchange 
amplitudes are ignored. Clearly those amplitudes are essential in
finding good predictions for both cross sections and analyzing powers.
Further discussion is given in the next sub-section.

\begin{figure}[h]
\begin{center}
\scalebox{0.4}{\includegraphics*{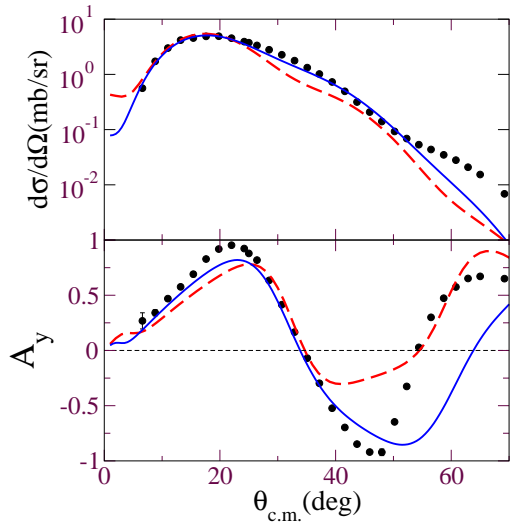}}
\end{center}
\caption[]{\label{Fig10}
The cross sections and analyzing powers for the inelastic scattering of
200 MeV protons exciting the $2^+$ 4.43 MeV state
of ${}^{12}$C. The solid curves are the complete results
and the dashed curves are those found when the medium effects
in the effective $NN$ interaction  are omitted.}
\end{figure}
The quality of the structure and the import of using medium modified
effective $NN$ interactions are confirmed by the results obtained for 
the inelastic scattering of 200 MeV protons exciting the 
the $2^+$ (4.43 MeV) state of ${}^{12}$C. Data for this reaction are
compared with microscopic DWA model results in Fig.~\ref{Fig10}.
Comparison of the two results with the data demonstrates
the important role played by medium effects in the $NN$ interaction
in this process. The variation in shapes of results for scattering angles
above 30$^\circ$ in the center of mass indicates that.
In addition the strength of the scattering is well predicted by the 
$(0+2)\hbar\omega$ structure model. When a simple $0\hbar\omega$ spectroscopy
was used, this cross section was a factor of four too weak in
comparison with the data~\cite{Am00} requiring upward scaling 
equivalent to an effective charge of $0.5e$ to match measurement.

\subsubsection{Effects of the non-local exchange amplitudes}

As the non-local
nature of the optical potentials formed by $g$-folding can be inconvenient,
in the past the exchange terms from which those non-localities arise
either have been ignored or approximated by a local potential.  While
the latter approach is the more sensible, we consider what happens
by ignoring the exchange terms and so retaining only the ``direct'' potentials;
the leading terms of Eq.~(\ref{NonSE}) i.e. those formed 
by folding with just the nuclear density and not the full density matrices.
By comparing complete with pure direct calculation results,
the influence of the exchange terms can be defined.

In the right-most panel of Fig.~\ref{Fig9}, the differential cross section and 
analyzing power data (200 MeV elastic proton scattering) are compared with 
the results of calculations made with the exchange amplitudes included,
without any simplification, and with those exchange amplitudes totally
excluded. The results are shown by the solid and dot-dashed curves
respectively.  The difference between 
the two sets of results is as noteworthy as the quality of agreement 
between the full calculation results and the data.  Such disparity between 
results with and without the exchange amplitudes
persist over a wide range of energies~\cite{Am00}.
Of particular note is that the direct and exchange contributions
destructively interfere to produce the final result. Also the two
amplitudes lead to very different momentum-transfer effects in order
that the cross-section results end up in the agreement with data found.

\subsubsection{Neutron scattering and effects of exchange amplitudes}

\begin{figure}[h]
\begin{center}
\scalebox{0.37}{\includegraphics*{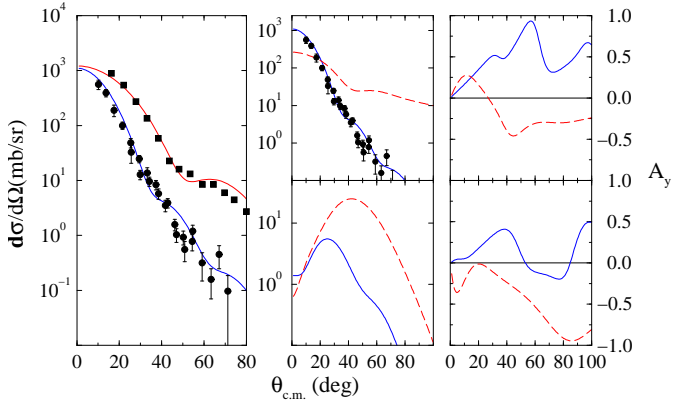}}
\end{center}
\caption[]{\label{Fig11} In the left panel are the cross sections for 
40.3 MeV and for 95 MeV neutrons scattered elastically from ${}^{12}$C.
In the middle panel are the cross sections for elastic (top) and inelastic
scattering (bottom) found with (solid) and without (dashed) the 
knock-out amplitudes.
The right panel are the analyzing powers associated with them.
Details are given in the text.}
\end{figure}

The cross section from the elastic scattering of 95 MeV neutrons from
${}^{12}$C has been measured recently~\cite{Kl03}  and these data are shown 
by the filled circles in the left panel of Fig.~\ref{Fig11}.  Older data taken 
at 40.3 MeV~\cite{Wi86} are shown therein by the filled squares. 
The middle and left panels show the results of 95 MeV neutron scattering 
elastic (top) and inelastic,
to the $2^+$ (4.43 MeV) state, (bottom) cross section
and analyzing powers respectively.
The solid curves depict the results obtained using the complete 
$g$-folding model with WS wave functions and orbit occupancies 
from the $(0+2)\hbar\omega$ shell model. The dashed curves depict
what result on omitting the exchange amplitudes.

At both energies, treating the exchange terms without approximation to get 
the optical potentials gave cross sections in very good agreement with the 
data.  The contrast of these results
with those given by the calculations made omitting the exchange amplitudes
is dramatic. The structure of the cross
sections and analyzing powers are radically changed as is the predicted 
magnitude of the inelastic excitation. These variations are similar to those
noted above with the (elastic) scattering of 200 MeV protons.

Of particular import is that, at all three energies considered, the role of 
the exchange (non-locality) in evaluations of both the 
elastic and inelastic-scattering cross sections is of destructive 
interference. Such interference makes representation of the non-local
interactions by an equivalent local interaction suspect. Also
the effects in the analyzing powers are dramatic
and it must be remembered that those observables involve the cross section
as a denominator. A fit to any analyzing power then without first having
a good cross section prediction is but fortuitous.


\section{Conclusions}

The origins and effects of non-localities in the $NA$ optical potential 
have been studied with separate methods of data analysis most appropriate 
for low and for medium projectile energies. The scattering of nucleons 
from ${}^{12}$C have been considered.

For low energies (0 to $\sim 6$ MeV) the MCAS method was used to describe 
$n$+${}^{12}$C scattering. With that theory
the origins of non-localities in optical potentials due to antisymmetry 
of projectiles with target nucleons and the dynamic polarization
potential representation of channel coupling were detailed. 
A collective model was used to specify the matrix of interaction potentials
that were the basic input to the approach.
The resulting DPP is strongly non-local with also strong $l$-dependence.
At the energies studied the dominant term is that of 
$0s_{\frac{1}{2}}$ wave interaction
whose character 
varies markedly to carry the resonance features of the full coupled-channel
 results.

At medium energies, the $g$-folding method with no-core, large space, 
Shell model wave
functions and the Melbourne force, a complex effective $NN$ interaction 
that is density and energy dependent, was used.
That approach attributes the effective channel coupling to be 
with an effective infinity of
target states in which one or more nucleons are in the continuum.
It is equivalent to using the KMT theory but with the effective 
interaction between
the projectile and each and every target nucleon being modified by the medium.
The cross sections obtained when compared with data, established
that one cannot ignore either medium modification of the $NN$ 
interaction or the exchange (knock-out) contributions
in forming optical potentials. 
That is observed strongly also when the DWA is used to evaluate observables 
from inelastic scattering.
The complete $g$-folding process makes the optical 
potential complex, energy dependent, and non-local. 
But, as the $g$-matrices are 
strongly medium dependent, the non-local attributes of the optical potentials
themselves are
not easily displayed. Indeed in the DWBA98 code such are not explicitly
evaluated in finding solutions of the integro-differential equations
from which phase shifts are specified~\cite{Am00}.
The import of that
non-locality is evident however in the comparisons made with data at
many energies and with and without the exchange terms included in the
calculations of the cross sections and analyzing powers.  Clearly, any 
localization of the non-local potential is approximating a most important
factor in data analysis.

\section*{acknowledgments}
This research was supported by the Italian MIUR-PRIN Project      
``Fisica Teorica del Nucleo e dei Sistemi a Pi\`u Corpi'', by the 
Natural Sciences and Engineering Research Council (NSERC), Canada, 
and by the National Research Foundation of South Africa.


\begin{thebibliography}{0}

\bibitem{Am00}
K. Amos, P. J. Dortmans, H.~V. von Geramb, S. Karataglidis, and 
J. Raynal, Adv. in Nucl. Phys. \textbf{25}, (2000);
and references cited therein.

\bibitem{Fe92}
H. Feshbach, {\it Theoretical Nuclear Physics: Nuclear Reactions},
(Wiley, 1992).

\bibitem{Pe62}
F. G. Perey and B. Buck, Nucl. Phys. \textbf{32}, (1962) 353.

\bibitem{Ca05}
L. Canton, G. Pisent, J. P. Svenne, D. van der Knijff, K. Amos,
 and S. Karataglidis, Phys. Rev. Lett. \textbf{94}, (2005) 122503.


\bibitem{Ke59}
A. K. Kerman, H. McManus, and R. M. Thaler, Ann. Phys. (N.Y.)
\textbf{8}, (1959) 551.

\bibitem{Ge75}
H. V. von Geramb, K. Amos, R. Sprickmann, K. T.
Kn{\"o}pfle, M. Rogge, D. Ingham, and C. Mayer-B{\"o}ricke,
Phys. Rev. C \textbf{12}, (1975) 1697.

\bibitem{Am05}
K. Amos, S. Karataglidis, D. van der Knijff, L. Canton, G. Pisent,
 and J. P. Svenne, Phys. Rev. C \textbf{72}, (2005) 065604.

\bibitem{Am03}
K. Amos, L. Canton, G. Pisent, J. P. Svenne, and D. van der Knijff,
Nucl. Phys. \textbf{A728}, (2003) 65.

\bibitem{Pi05}
G. Pisent, J. P. Svenne, L. Canton, K. Amos, S. Karataglidis, and 
D. van der Knijff, Phys. Rev. C \textbf{72}, (2005) 014601.

\bibitem{Ca91}
G. Cattapan, L. Canton, and G. Pisent, {\it Phys. Rev. C}
\textbf{43}, (1991) 1395.

\bibitem{Ca90}
G. Cattapan, L. Canton, and G. Pisent, {\it Phys. Lett.} 
\textbf{B 240}, (1990) 1.

\bibitem{Ca91a}
L. Canton, Y. Hahn, and G. Pisent, {\it Phys. Rev. C}
\textbf{43}, (1991) 2441.

\bibitem{Be63}
H. A. Bethe, B. H. Brandow, and A. G. Petschek, Phys. Rev.
\textbf{129}, (1963) 225.

\bibitem{Ar96}
H. F. Arellano, F. A. Brieva, M. Sander, and H. V. von Geramb,
Phys. Rev. C \textbf{54}, (1996) 2570.

\bibitem{Ra98}
J. Raynal, {\it computer program DWBA98}, NEA 1209/05, (1998).

\bibitem{Pe93}
S. Pearlman, {\it ENDF/HE-VI Mat-625}, BNL-48035, (1993).

\bibitem{IA05}
{\it Computer Index of Neutron Data} ({\bf CINDA})
IAEA - NDS database (2005).

\bibitem{Sv06}
J. P. Svenne, K. Amos, S. Karataglidis, D. van der Knijff, L. Canton,
 and G. Pisent, Phys. Rev. C \textbf{73}, (2006) 027601.

\bibitem{Ka95}
S. Karataglidis, P. J. Dortmans, K. Amos, and R. de Swiniarski,
Phys. Rev. C \textbf{52}, (1995) 861. 

\bibitem{Ja72}
J. A. Jansen, R. Th. Peerdeman, and C. de Vries, Nucl. Phys.
\textbf{A188}, (1972) 337.

\bibitem{Si70}
I. Sick and J. S. McCarthy, Nucl. Phys. \textbf{A150}, (1970) 631.

\bibitem{Na71}
A. Nakada, Y. Torizuka, and Y. Horikawa, Phys. Rev. Lett.
\textbf{27}, (1971) 745, 1102.  

\bibitem{Fl78}
J. B. Flanz, R. S. Hicks, R. A. Lindgren, G. A. Peterson, A. Hotta, 
B. Parker, and R. C. York, Phys. Rev. Lett. \textbf{41}, 
(1978) 1642.

\bibitem{Co82}
J. R. Comfort, G. L. Moake, C. C. Foster, P. Schwandt, and W. G. Love,
Phys. Rev. C \textbf{26}, (1982) 1800.

\bibitem{Co65}
S. Cohen and D. Kurath, {\it Nucl. Phys.}
\textbf{73}, (1965) 1.


\bibitem{Kl03}
J. Klug {\it et al.} Phys. Rev. C \textbf{67}, (2003) 031601.

\bibitem{Wi86}
J. S. Winfield, S. M. Austin, R. P. DeVito, U. E. Berg,
Z. Chen, and W. S. Sterrenburg, Phys. Rev. C \textbf{33}, (1986) 1.

\end{thebibliography}
\end{document}